\documentclass{ws-procs9x6}
\usepackage{braket}

\begin{document}

\title{
Microscopic Properties of Quantum Annealing\\
-- Application to Fully Frustrated Ising Systems --}

\author{Shu Tanaka
\footnote{Present address: Department of Chemistry, University of Tokyo,
7-3-1 Hongo, Bunkyo-ku, Tokyo 113-0033, Japan}
}

\address{
Research Center for Quantum Computing, 
Interdisciplinary Graduate School of Science and Engineering, Kinki University, 
3-4-1 Kowakae, Higashi-Osaka, Osaka 577-8502, Japan\\
E-mail: shu-t@chem.s.u-tokyo.ac.jp
}

\begin{abstract}
In this paper we show quantum fluctuation effect of fully frustrated Ising spin systems.
Quantum annealing has been expected to be an efficient method to find ground state of optimization problems.
However it is not clear when to use the quantum annealing.
In order to clarify when the quantum annealing works well, we have to study microscopic properties of quantum annealing.
In fully frustrated Ising spin systems, there are macroscopically degenerated ground states.
When we apply quantum annealing to fully frustrated systems, we cannot obtain each ground state with the same probability.
This nature is consistent with ``order by disorder'' which is well-known mechanism in frustrated systems.
\end{abstract}


\bodymatter

\section{Introduction}

Quantum information processing has been expected to be able to solve difficult problems which are not tractable on classical computer\cite{Shor-1994,Nielsen-2000,Nakahara-2008}.
Then a number of researchers have studied how to realize a quantum computer and quantum algorithm itself.
Study on quantum information from a viewpoint of statistical physics came into the world about a decade ago.
This is called quantum annealing\cite{Kadowaki-1998,Farhi-2001,Santoro-2002,Kadowaki-2002,Charkrabarti-book1,Das-2008,Charkrabarti-book2}, in other words, quantum adiabatic evolution.
In this paper we review on microscopic properties of quantum annealing with providing a specific example of Ising spin systems.

The Ising model has been regarded as a standard model in statistical physics since it can be represented nature of phase transition and a couple of physical phenomena in real magnetic materials.
The Hamiltonian of the Ising model with random bonds is given as
\begin{eqnarray}
 \label{eq:IsingHam}
 {\cal H} = -\sum_{\langle i,j \rangle} J_{ij} \sigma_i^z \sigma_j^z,
  \qquad \sigma_i^z = \pm 1,
\end{eqnarray}
where $\langle i,j \rangle$ represents pairs of sites on the given graph.
Since the spin variable $\sigma_i^z$ takes $\pm 1$, this variable $\sigma_i^z$ obviously can be regarded as a ``bit''.
Then the Ising model has been adopted not only for magnetic materials but also for information science.
For example, traveling salesman problems and neural network can be mapped onto the random bond Ising model.
These problems are categorized into optimization problems.
It is difficult to find the best solution of optimization problems, since the number of candidate of solutions increases exponentially as the number of elements increases.
Then we have to use engineered method to search the best solution of optimization problems.
One of the methods is Monte Carlo simulation.
However if we use Monte Carlo simulation, we face on difficulty of search the best solution of optimization problem, since energy landscape of optimization problems is complicated in general.
In order to avoid such a difficulty, a number of researchers have improved the standard Monte Carlo simulation.

\begin{figure}[t]
 \begin{center}
  \psfig{file=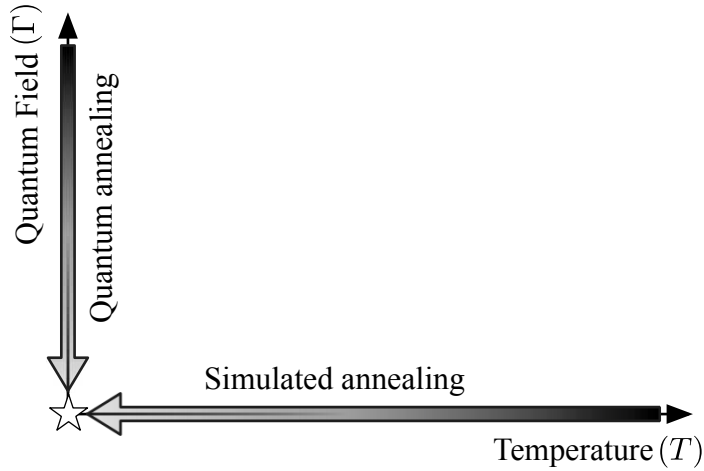,scale=1}
 \end{center}
 \caption{
 Schematic picture of simulated annealing and quantum annealing.
 The star denotes the position of $T=0$ and $\Gamma=0$ which is the target position.}
 \label{fig:saqa}
\end{figure}

One of the important methods is simulated annealing which was first proposed by Kirkpatrick\cite{Kirkpatrick-1983,Kirkpatrick-1984}.
Roughly speaking, the simulated annealing is an optimization technique by decreasing temperature {\it i.e.} thermal fluctuation (Fig.~\ref{fig:saqa}).
About fifteen years later since the publication of Kirkpartick's pioneering work, an alternative method of simulated annealing was proposed by Kadowaki and Nishimori\cite{Kadowaki-1998}.
In order to obtain the ground state, we decrease quantum field as substitute for temperature (Fig.~\ref{fig:saqa}).
The method is called quantum annealing.
The quantum annealing is expected to be an efficient method to obtain the better solution of optimization problems.
Actually, the performance of quantum annealing has been demonstrated by a number of researchers\cite{Tanaka-2002,shu-t-2007a,Matsuda-2009,Kurihara-2009,Sato-2009,shu-t-2009a,shu-t-2010a,shu-t-2010b,shu-t-2010d}.

However the solution obtained by the quantum annealing is worse than that obtained by the simulated annealing for some cases\cite{Battaglia-2005}.
It is not clear when to use and when not to use the quantum annealing.
Then we should study microscopic mechanism of quantum annealing in order to understand why the quantum annealing works well for many cases.
In the next section, we will review how to implement standard Monte Carlo simulation, simulated annealing, and quantum annealing.
In Section 3, we will consider a quantum fluctuation effect of frustrated Ising spin systems.
Finally we will summarize the microscopic mechanism of quantum annealing and show future perspective.

\section{Computation Method}

Many optimization problems can be represented by random Ising models as stated in the previous section.
The ground state of the random Ising spin system corresponds to the best solution of the given optimization problem.
There can be many strategy to find a ground state of these systems.
We often use Monte Carlo method as one of the most general algorithm.
In the first part of this section, we will review on Monte Carlo simulation and simulated annealing. 
Next we will introduce quantum annealing which is regarded as an alternative for the simulated annealing.

Here we consider the Hamiltonian given by Eq.~(\ref{eq:IsingHam}).
Algorithm of Monte Carlo simulation can be summarized as follows:
\begin{description}
 \item[step 1] By using random number generator, we prepare a random state as the initial state.
 \item[step 2] We select a site at random and calculate an effective field.
	    The effective field is defined as
	    \begin{eqnarray}
	     h_{\rm eff}^{(i)} = \sum_{j \in {\rm n.n.\, of }\, i} J_{ij} \sigma_j^z,
	    \end{eqnarray}
	    where $i$ denotes the label of the selected site.
	    The summation takes over the nearest neighbor (n.n.) of the $i$-th site.
 \item[step 3] We flip the spin of the $i$-th site according to some kind of transition probability.
	    How to choose the transition probability will be given below.
 \item[step 4] We repeat step 2 and step 3 for a long time.
\end{description}
There can be a couple of definitions of transition probability in step 3.
For example, heat bath method and Metropolis method are very famous and often adopted.
Transition probability in heat bath method and in Metropolis method are given as
\begin{eqnarray}
 &&w_{\rm HB}(\sigma_i^z \to -\sigma_i^z) =
  \frac{{\rm e}^{-\beta h_{\rm eff}^{(i)}\sigma_i^z}}{{\rm e}^{\beta h_{\rm eff}^{(i)}\sigma_i^z}+{\rm e}^{-\beta h_{\rm eff}^{(i)}\sigma_i^z}},\\
 &&w_{\rm MP}(\sigma_i^z \to -\sigma_i^z) = 
  \begin{cases}
   1 &(\sigma_i^z h_{\rm eff}^{(i)}<0),\\
   {\rm e}^{-\beta \sigma_i^z h_{\rm eff}^{(i)}}
   & (\sigma_i^z h_{\rm eff}^{(i)} \ge 0),
  \end{cases}
\end{eqnarray}
Both two transition probabilities obey the detailed balance condition.
Very recently, an efficient algorithm was proposed by Suwa and Todo\cite{Suwa-2010}.
Their algorithm is not based on the detailed balance condition but on balance condition only.
It helps us to obtain in a shorter time than conventional method, although it depends on problem.

At high temperature, state can be changed easily since the canonical distribution is almost flat.
However it is generally difficult to change state at low temperature because of complicated energy landscape and relatively high energy barrier.
Then we can avoid partly such a difficulty by decreasing the temperature gradually.
This is called simulated annealing.
It has been widely adopted for many cases.

Next we review how to implement Monte Carlo simulation for random Ising systems with transverse field.
The Hamiltonian is given as
\begin{eqnarray}
 &&{\cal H} = {\cal H}_{\rm c} + {\cal H}_{\rm q},\\
 &&{\cal H}_{\rm c} := - \sum_{\langle i,j \rangle} J_{ij} \sigma_i^z \sigma_j^z,
  \qquad
  {\cal H}_{\rm q} := -\Gamma \sum_i \sigma_i^x,
\end{eqnarray}
where $\sigma_i^x$ and $\sigma_i^z$ represent the $x$-component and $z$-component of the Pauli matrix at the $i$-th site, respectively.
Here we adopt transverse field as a quantum term.
For small systems, we can obtain equilibrium physical quantities since we can calculate the Boltzmann weight ${\rm e}^{-\beta {\cal H}}$ by using exact diagonalization.
However it is difficult to calculate the Boltzmann weight ${\rm e}^{-\beta {\cal H}}$ for relatively large system because of limitation of memory.
Then, in general, we have to adopt other strategy for obtaining the equilibrium properties of large-scale quantum systems.
Nowadays the quantum Monte Carlo simulation is regarded as an efficient algorithm.
In order to use Monte Carlo method, we have to calculate equilibrium probability ${\rm e}^{-\beta {\cal H}}$ as mentioned above.
Here we calculate partition function of the original quantum system ${\cal H}$ by applying the Trotter-Suzuki decomposition\cite{Trotter-1959,Suzuki-1976}.
The partition function is given by
\begin{eqnarray}
 Z = {\rm Tr}\, {\rm e}^{-\beta {\cal H}}
  = {\rm Tr}\, {\rm e}^{-\beta ({\cal H}_{\rm c} + {\cal H}_{\rm q})}
  = \sum_{\Sigma} \braket{\Sigma|{\rm e}^{-\beta ({\cal H}_{\rm c} + {\cal H}_{\rm q})}|\Sigma}.
\end{eqnarray}
By using the following relation
\begin{eqnarray}
  {\rm e}^{-\frac{1}{m}\beta({\cal H}_{\rm c}+ {\cal H}_{\rm q})}
  = {\rm e}^{-\frac{1}{m}\beta{\cal H}_{\rm c}}
  {\rm e}^{-\frac{1}{m}\beta{\cal H}_{\rm q}}
  + {\cal O}\left( \left( \frac{\beta}{m}\right)^2\right),
\end{eqnarray}
we obtain 
\begin{eqnarray}
 \nonumber
 Z = \sum_{\{\tilde{\sigma}_{i,k},\tilde{\sigma}_{i,k}'=\pm 1\}}&&
  \braket{\Sigma_1|{\rm e}^{-\frac{\beta{\cal H}_{\rm c}}{m}}|\Sigma_1'}
  \braket{\Sigma_1'|{\rm e}^{-\frac{\beta{\cal H}_{\rm q}}{m}}|\Sigma_2}\\
\nonumber  
&&\times
  \braket{\Sigma_2|{\rm e}^{-\frac{\beta{\cal H}_{\rm c}}{m}}|\Sigma_2'}
  \braket{\Sigma_2'|{\rm e}^{-\frac{\beta{\cal H}_{\rm q}}{m}}|\Sigma_3}\\
\nonumber  
&&\times \cdots\\
 \label{eq:partitionfunction}
  &&\times
  \braket{\Sigma_m|{\rm e}^{-\frac{\beta{\cal H}_{\rm c}}{m}}|\Sigma_m'}
  \braket{\Sigma_m'|{\rm e}^{-\frac{\beta{\cal H}_{\rm q}}{m}}|\Sigma_1},
\end{eqnarray}
where $\ket{\Sigma_k}$ expresses direct product such as
\begin{eqnarray}
 \ket{\Sigma_k} := \ket{\tilde{\sigma}_{1,k}} \otimes 
  \ket{\tilde{\sigma}_{2,k}} \otimes
  \cdots
  \otimes
  \ket{\tilde{\sigma}_{N,k}},
\end{eqnarray}
where $N$ represents the number of sites and $k$ denotes the position along the Trotter axis.
The first subscript of $\tilde{\sigma}$ is the position in the real space.
Since ${\cal H}_{\rm c}$ is a diagonal matrix, then we obtain
\begin{eqnarray}
 \braket{\Sigma_k|{\rm e}^{-\frac{\beta{\cal H}_{\rm c}}{m}}|\Sigma_k'}
  = {\rm e}^{\frac{\beta}{m} \sum_{\langle i,j \rangle} J_{ij} \tilde{\sigma}_{i,k} \tilde{\sigma}_{j,k}}
  \prod_{i=1}^N \delta_{\tilde{\sigma}_{i,k},\tilde{\sigma}_{i,k}'}.
\end{eqnarray}
On the other hand, we can calculate $\braket{\Sigma_k'|{\rm e}^{-\frac{\beta {\cal H}_{\rm q}}{m}}|\Sigma_{k+1}}$ as follows:
\begin{eqnarray}
 \braket{\Sigma_k'|{\rm e}^{-\frac{\beta{\cal H}_{\rm q}}{m}}|\Sigma_{k+1}}
  = \left[
     \frac{1}{2} \sinh \left( \frac{2\beta\Gamma}{m}\right)
    \right]^{\frac{1}{2}}
  {\rm e}^{\left[ \frac{1}{2} \log \coth \left( \frac{\beta\Gamma}{m} \sum_{i=1}^N \tilde{\sigma}_{i,k} \tilde{\sigma}_{i,k+1}'\right) \right]}
\end{eqnarray}
The partition function of the original Hamiltonian ${\cal H}$ is expressed as
\begin{eqnarray}
\nonumber
 Z =&& \lim_{m \to \infty}
  \left[
   \frac{1}{2} \sinh \left( \frac{2\beta\Gamma}{m}\right)
  \right]^{\frac{N}{2}}\\
 \nonumber
  &&\times 
   \sum_{\{\tilde{\sigma}_{i,k}=\pm 1\}}
   {\rm e}^{
  \left[
   \sum_{\langle i,j \rangle} \sum_{k=1}^m
   \left(
    \frac{\beta J_{ij}}{m} \tilde{\sigma}_{i,k} \tilde{\sigma}_{j,k}
   \right)
  + \frac{1}{2} \sum_{i=1}^N \sum_{k=1}^m
  \log \coth \left( \frac{\beta \Gamma}{m}\right)
  \tilde{\sigma}_{i,k} \tilde{\sigma}_{i,k+1}
  \right]}.
\end{eqnarray}
Then the equivalent classical effective Hamiltonian ${\cal H}_{\rm e}$ can be written as
\begin{eqnarray}
 {\cal H}_{\rm e} = - \sum_{\langle i,j \rangle} \sum_{k=1}^m \frac{J_{ij}}{m} \tilde{\sigma}_{i,k}^z \tilde{\sigma}_{j,k}^z 
  - \frac{1}{2\beta} \log \coth \left( \frac{\beta\Gamma}{m} \right)
  \sum_{i=1}^N\sum_{k=1}^m \tilde{\sigma}_{i,k} \tilde{\sigma}_{i,k+1}.
\end{eqnarray}
Note that $\tilde{\sigma}_{i,m+1}=\tilde{\sigma}_{i,1}$ because of Eq.~(\ref{eq:partitionfunction}).
We can obtain equilibrium properties of the original quantum system ${\cal H}$ by calculating that of the classical effective Hamiltonian ${\cal H}_{\rm e}$.
We can perform the quantum annealing by decreasing transverse field $\Gamma$.

\section{Quantum Fluctuation Effect of Frustrated Ising Systems}

In this section we consider quantum fluctuation effect of frustrated Ising systems comparing with thermal fluctuation effect.
Although there are a couple of types of frustrated systems such as geometric frustrated systems and random systems, we focus on regularly frustrated systems in this paper.

\begin{figure}[b]
 \begin{center}
  \psfig{file=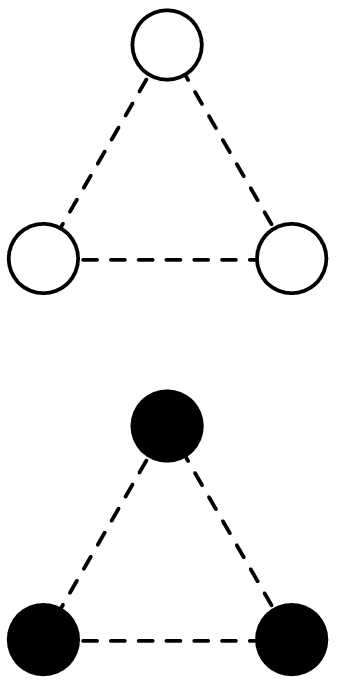,scale=0.3}
  \hspace{1.5cm}
  \psfig{file=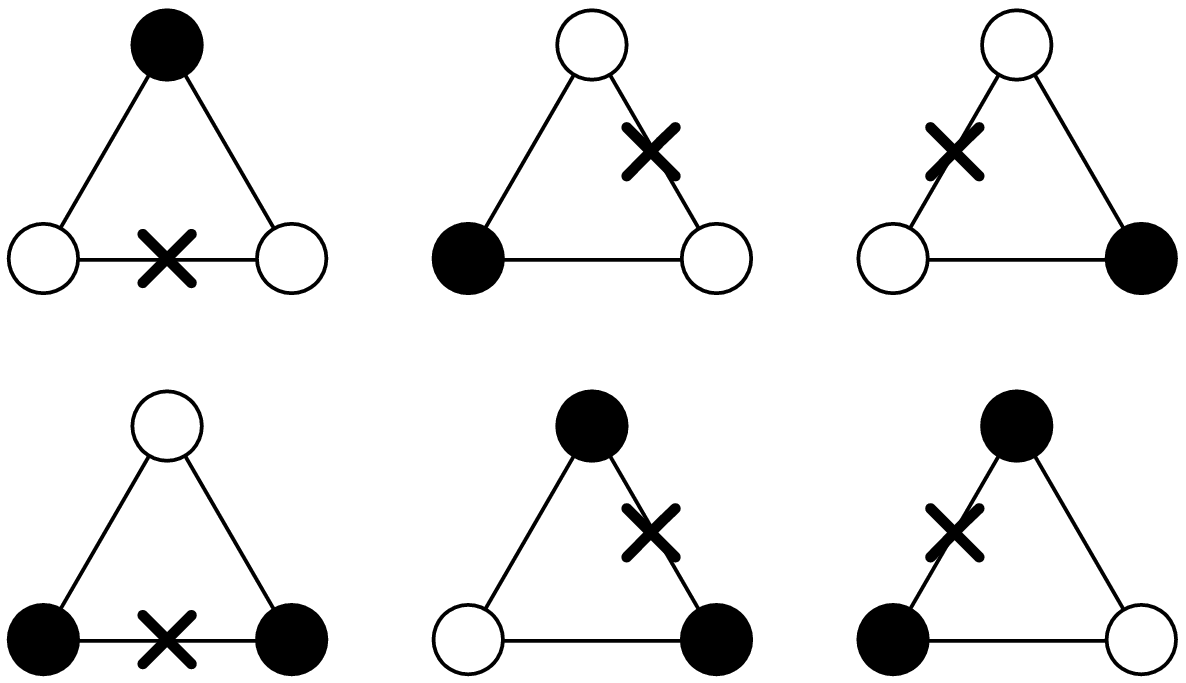,scale=0.3}
 \end{center}
 \caption{
 The white and black circles denote $+$ spins and $-$ spins, respectively.
 (Left panel) The ground states of the ferromagnetic triangle cluster.
 (Right panel) The ground states of the antiferromagnetic triangle cluster.}
 \label{fig:3spins_gs}
\end{figure}

Figure \ref{fig:3spins_gs} shows all ground states of three spin systems.
The dotted lines and the solid lines in Fig.~\ref{fig:3spins_gs} represent ferromagnetic interactions and antiferromagnetic interactions, respectively.
The white and black circles in Fig.~\ref{fig:3spins_gs} indicate $+$ spins and $-$ spins, respectively.
As shown in Fig.~\ref{fig:3spins_gs}, there are six ground states for antiferromagnetic case.
The crosses of the right panel in Fig.~\ref{fig:3spins_gs} depict energetically unfavorable interactions.
In general, there are many degenerated ground states in frustrated systems because of such unfavorable interactions.

\begin{figure}[t]
 \begin{center}
  \psfig{file=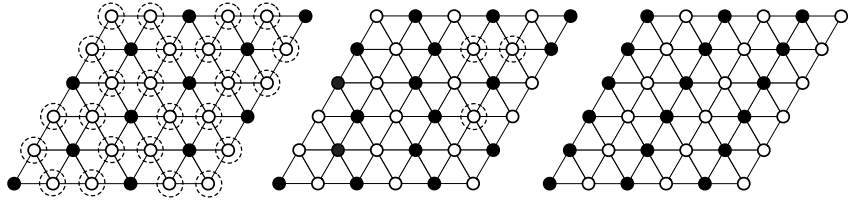,scale=0.65}
 \end{center}
  \caption{
  Ground states of antiferromagnetic Ising system on triangular lattice.
  The dotted circles denote free spins where the effective field is zero.
 The left panel shows one of the maximum free spin states.
  }
  \label{fig:triangle_gs}
\end{figure}

\begin{figure}[t]
 \begin{center}
  \psfig{file=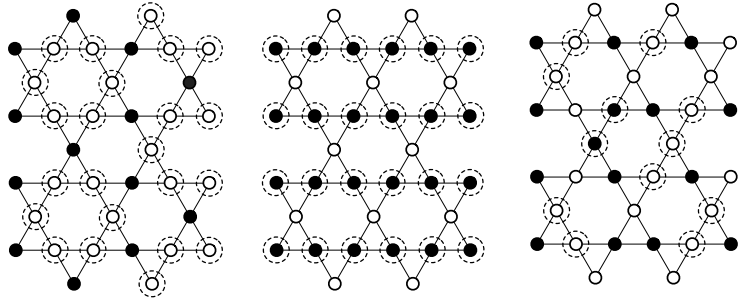,scale=0.65}
 \end{center}
 \caption{
 Ground states of antiferromagnetic Ising system on kagome lattice.
 The dotted circles denote free spins where the effective field is zero.
 The left and middle panel show maximum free spin states.
 }
 \label{fig:kagome_gs}
\end{figure}

One of regularly frustrated systems is antiferromagnetic Ising model on triangular lattice.
Figure~\ref{fig:triangle_gs} takes three of many ground states of this system.
Figure~\ref{fig:kagome_gs} also takes three of many ground states of antiferromagnetic Ising model on kagome lattice as an another example.
In these models, as the number of spins $N$ increases, the number of ground states increases exponentially.

Our motivation is as follows:
Suppose we consider systems where there are many ground states.
When we apply quantum annealing to such systems, which states are selected at the final time? 
What are similarities and differences between simulated annealing and quantum annealing? 
In order to consider them, we study quantum fluctuation effect of regularly frustrated Ising spin systems.

Before we study quantum fluctuation effect, we consider thermal fluctuation effect for comparison.
We apply simulated annealing to regularly frustrated Ising systems.
Suppose the cooling speed is set to be very slow from high temperature to $T=0+$.
Then we can obtain each ground state with the same probability.
This is a reasonable nature from a viewpoint of the principle of statistical physics.
If the internal energy in the $i$-th state and that in the $j$-th state are the same, probabilities of both states are the same.
This is called the principle of equal weight.
Then such nature appears despite details of model such as lattice structure.

On the other hand, situation is drastically changed when we apply quantum annealing to regularly frustrated Ising systems even if we decrease transverse field very slow.
The dotted circles in Fig.~\ref{fig:triangle_gs} and Fig.~\ref{fig:kagome_gs} represent spins where the effective field is zero.
The states with the highest appearance probability corresponds to the maximum free spin states.
As the number of free spins decreases, the appearance probability decreases at least in large number of free spin region.
This result is consistent with the Villain model case which is a typical example of regularly fully frustrated Ising systems as well as antiferromagnetic Ising model on triangular lattice and kagome lattice.

\begin{figure}[t]
 \begin{center}
  \psfig{file=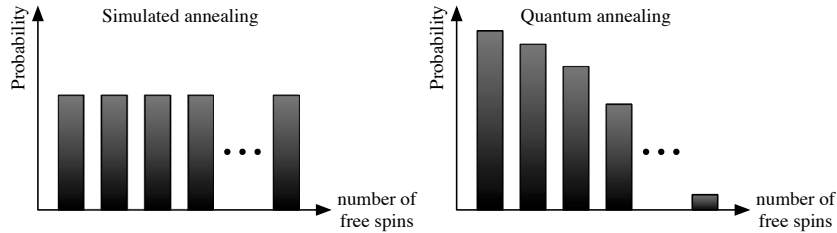,scale=0.6}
 \end{center}
 \caption{
 Schematic picture of probability distribution at the final time for simulated annealing (left panel) and quantum annealing (right panel) for adiabatic limit.
 }
 \label{fig:probdis}
\end{figure}

Finally, Fig.~\ref{fig:probdis} represents a schematic picture of probability distribution at the final time for simulated annealing and quantum annealing for adiabatic limit.

\section{Conclusion}

In this paper, we review implementation method of Monte Carlo simulation of classical Ising models and quantum Ising models.
We also show microscopic properties of quantum annealing for fully frustrated Ising spin systems comparing with that of simulated annealing.
When we apply quantum annealing to fully frustrated Ising systems, we cannot obtain each ground state with the same probability.
In contrast, we can obtain each ground state with the same probability by simulated annealing because of the principle of statistical physics.
In this paper we consider adiabatic limit of quantum annealing.
We have to study microscopic properties of quantum annealing beyond adiabatic limit keeping practical situations in our mind.

In this study, we focus on transverse field as a quantum fluctuation.
It is not necessary that we restrict quantum fluctuation to transverse field.
Moreover we can adopt a new type of fluctuation such as invisible fluctuation which is first proposed by Tamura {\it et al.}\cite{Tamura-2010,Tanaka-2010,Tanaka-2011a,Tanaka-2011b}
In order to improve original simulated annealing, we have to study fluctuation effect of random systems in a general way.

Quantum annealing, in other words, quantum adiabatic evolution was born in frontier between quantum information science and quantum statistical physics.
Study on quantum annealing is quite primitive and will have a favorable influence on both fields\cite{Tanaka-inpre}.

The author is grateful to Masaki Hirano, Kenichi Kurihara, Yoshiki Matsuda, Seiji Miyashita, Hiroshi Nakagawa, and Issei Sato for fruitful discussion.
The author also thanks Ryo Tamura for critical reading.
The author is partly supported by Grant-in-Aid for Young Scientists Start-up (21840021) from the JSPS, MEXT Grant-in-Aid for Scientific Research (B) (22340111), and the ``Open Research Center'' Project for Private Universities: matching fund subsidy from MEXT.
The computation in the present work was performed on computers at the Supercomputer Center, Institute for Solid State Physics.
\bibliographystyle{ws-procs9x6}
\bibliography{ws-pro-sample}

\end{document}